\documentclass[twocolumn,10pt,pra,aps,superscriptaddress,showpacs,amsmath,amssymb]{revtex4-1}

\usepackage[english]{babel}
\usepackage{booktabs}

\usepackage{amsmath}
\usepackage{amssymb}
\usepackage{graphicx}
\usepackage{array}
\usepackage[colorlinks=true, allcolors=blue]{hyperref}
\usepackage{algorithm}
\usepackage{algpseudocode}

\begin{document}
\title{Circuit-level benchmarks of GKP-concatenated qLDPC Codes}
\author{Yuan Yao}
\thanks{These authors contributed equally to this work.}
\affiliation{Center on Frontiers of Computing Studies, School of Computer Science, Peking University, Beijing 100871, China}
\author{Ruipeng Xing}
\thanks{These authors contributed equally to this work.}
\affiliation{Center on Frontiers of Computing Studies, School of Computer Science, Peking University, Beijing 100871, China}%
\author{Jianshuo Gao}
\affiliation{Center on Frontiers of Computing Studies, School of Computer Science, Peking University, Beijing 100871, China}
\author{Xiao Yuan}
\affiliation{Center on Frontiers of Computing Studies, School of Computer Science, Peking University, Beijing 100871, China}

\begin{abstract}
Scalable fault-tolerant quantum error correction based on Gottesman--Kitaev--Preskill (GKP) codes requires finite-rate outer codes that can exploit analog GKP information under realistic circuit noise. However, quantum low-density parity-check (qLDPC) outer codes have not been systematically compared as candidates for circuit-level GKP concatenation. Here, we benchmark BB and tricycle codes using a unified simulation framework that progresses from code-capacity noise to repeated noisy syndrome extraction and schedule-resolved displacement propagation. Across all three noise models, analog-informed belief-propagation with ordered-statistics decoding (BP--OSD) yields higher finite-size crossing estimates than hard-decision decoding. In the full circuit model, the analog-informed crossings are \(\sigma_{\mathrm{th}}\simeq0.212\) for the selected BB sequence and \(0.142\) for the tricycle sequence, corresponding to squeezing requirements of approximately \(10.46\,\mathrm{dB}\) and \(13.94\,\mathrm{dB}\), respectively. These results show that analog GKP information continues to improve decoding under circuit-level noise. Under the common simulation assumptions used here, the selected BB sequence also yields a higher finite-size crossing estimate than the selected tricycle sequence.
\end{abstract}
\maketitle

\section{Introduction}

Quantum error correction (QEC) protects quantum information against decoherence, control errors, and measurement noise~\cite{shor1995scheme,steane1996error}. Most established fault-tolerant architectures encode information in discrete-variable qubits and use stabilizer measurements to identify Pauli errors~\cite{steane1996error,fowler2012surface}. Many hardware platforms, however, naturally provide bosonic modes whose dominant errors are small displacements in phase space. Bosonic codes exploit the larger Hilbert space of an oscillator to suppress these errors before they become logical faults~\cite{gottesman2001encoding,albert2018performance}.

Among bosonic encodings, the Gottesman--Kitaev--Preskill (GKP) code is particularly attractive because it encodes a logical qubit in a single oscillator while preserving analog information about displacement errors~\cite{gottesman2001encoding}. Experimental progress in superconducting cavities, trapped ions, and integrated photonics has made GKP-based quantum information processing increasingly concrete~\cite{campagneIbarcq2020quantum,sivak2023realtime,matsos2025universal,larsen2025integrated}. Breeding-based approaches to optical GKP-state preparation have also been developed using cat-state interference, Gaussian operations, and conditional measurements~\cite{weigand2018generating,takase2023synthesizer}. Complementing these protocols, an integrated photonic source based on Gaussian boson sampling and photon-number-resolving heralding has experimentally generated optical GKP states on a silicon-nitride platform~\cite{larsen2025integrated}. A more recent hardware-oriented proposal combines deterministic squeezed-cat preparation, cat breeding, and homodyne conditioning in a strongly coupled quantum-dot--cavity system to generate finite-energy GKP resource states~\cite{kamath2026generating}.

Despite this progress, a single finite-energy GKP layer does not provide scalable fault tolerance. Finite squeezing, imperfect state preparation, noisy operations, and measurement errors leave residual logical Pauli faults after GKP correction~\cite{gottesman2001encoding,noh2020fault}. Concatenation with an outer qubit code provides a systematic way to suppress these residual errors~\cite{fukui2018high,noh2020fault,noh2022lowoverhead}. The general principle dates back to Knill and Laflamme: an inner code modifies and suppresses the physical noise, while an outer code corrects the remaining logical faults~\cite{knill1996concatenated}. Recent studies of concatenated quantum Hamming codes have renewed interest in this approach by developing constant- or low-overhead fault-tolerant constructions, more detailed resource analyses, and improved hierarchical decoding methods~\cite{yamasaki2024time,yoshida2025concatenate,tamiya2025polylog,zhang2026bidirectional}.

The same concatenation principle extends naturally to bosonic systems, where a bosonic inner code is combined with an outer stabilizer code~\cite{fukui2018high,chamberland2022building}. More broadly, the hardware efficiency of bosonic encoding has been demonstrated experimentally with a binomially encoded microwave-cavity qubit whose logical lifetime was extended beyond the break-even point through repeated syndrome extraction and real-time feedback~\cite{ni2023beating}. In GKP-concatenated architectures, the inner layer converts continuous displacement noise into residual qubit-level errors and supplies reliability information, while the outer code uses global syndrome constraints to infer a recovery~\cite{fukui2018high,vuillot2019quantum,noh2020fault}. Previous proposals have combined GKP qubits with toric, surface, color, repetition, and bias-adapted outer codes~\cite{vuillot2019quantum,noh2020fault,zhang2021quantum,stafford2023biased,zhang2023concatenation,noh2022lowoverhead}. A passive-linear-optical measurement-based architecture has also been developed for implementing arbitrary GKP-concatenated codes, with numerical benchmarks reported for hyperbolic surface codes and BB codes~\cite{walshe2025linear}.

Although GKP concatenation requires additional state preparation, inner correction, homodyne measurement, and feedback, these operations reshape the noise presented to the outer code. The GKP layer converts small phase-space displacements into discrete logical errors and assigns different reliabilities to different correction outcomes. An analog-informed outer decoder can retain this information instead of reducing every outcome to an equally reliable binary decision~\cite{fukui2017analog,fukui2018high,vuillot2019quantum}. If this improved effective noise model compensates for the additional bosonic operations, the outer code may require less overhead to reach a target logical error rate~\cite{noh2022lowoverhead}.

qLDPC codes are particularly interesting outer-code candidates in this setting. Their sparse parity-check matrices lead to bounded- or low-weight stabilizer measurements and Tanner graphs compatible with message-passing decoders, while their nonvanishing rates offer a possible asymptotic overhead advantage over low-rate topological codes. BB and tricycle codes provide two structured qLDPC families with different parity-check geometries, stabilizer weights, and syndrome-extraction circuits. Comparing them within the same GKP framework therefore makes it possible to study how the choice of outer code interacts with analog information and circuit-level displacement propagation.

In this work, we develop a common architectural and numerical framework for GKP-concatenated BB and tricycle codes. The architecture separates inner GKP correction, outer-code syndrome extraction, analog-information processing, and classical decoding, while exposing the circuit ingredients relevant to noise propagation and resource bookkeeping. We then construct three compatible noise models with increasing circuit detail: an ideal code-capacity model, a statistical variance-aggregation model with repeated noisy syndrome measurements, and a schedule-resolved circuit model that propagates continuous displacements through outer-code syndrome extraction. In the final model, inner GKP correction is represented through its induced Pauli transition and modular outcome, while ancilla, gate, measurement, idle, and propagation effects in the outer syndrome-extraction circuit are modeled explicitly. Hard-decision and analog-informed BP--OSD decoding are evaluated on the same sampled noise realizations, isolating the contribution of GKP reliability information.

We apply this framework to selected finite-length BB and tricycle instances under common noise, circuit, and decoding assumptions. Analog-informed decoding improves the finite-size crossing estimates for both sequences in all three noise models. From code capacity to variance aggregation and the full circuit model, the analog-informed estimates are \(\sigma_{\mathrm{th}}\simeq(0.556,0.369,0.212)\) for the BB sequence and \((0.500,0.349,0.142)\) for the tricycle sequence. The studied BB sequence maintains higher analog-informed crossings across all three levels of modeling. Because the two sequences differ in block length, rate, and distance, these values are finite-length benchmarks for the selected instances rather than universal thresholds of the complete code families. Together, the results demonstrate that the benefit of analog GKP information persists under circuit-level noise and show how outer-code structure affects the performance of GKP-concatenated qLDPC architectures.

\section{GKP-Concatenated Processing Stack}
\label{sec:gkp_architecture}

A GKP-concatenated architecture combines mode-level bosonic correction with block-level stabilizer correction. The inner layer converts continuous displacements of individual oscillator modes into effective Pauli faults and produces analog information about the reliability of each conversion. The outer layer uses stabilizer constraints across multiple GKP modes to infer a global recovery. These two layers are connected by the syndrome-extraction circuit and the classical decoder, so their performance cannot be analyzed independently.

The complete processing stack consists of four components: inner GKP correction, outer-code syndrome extraction, analog-information processing, and global decoding. This organization identifies where continuous noise becomes a discrete fault, how errors propagate through the correction circuit, and which information is retained by the decoder. It also provides a common framework for comparing different outer codes under the same physical and decoding assumptions.

\subsection{Architecture and Information Flow}
\label{subsec:overall_architecture}

Let the outer code encode \(k_{\mathrm{out}}\) logical qubits into
\(n_{\mathrm{out}}\) physical qubits. In the concatenated construction, each
physical qubit of the outer code is replaced by one GKP-encoded oscillator. The
resulting code can be represented schematically as
\begin{equation}
    \mathcal{C}
    =
    \mathcal{C}_{\mathrm{outer}}
    \circ
    \mathcal{C}_{\mathrm{GKP}} .
    \label{eq:concatenated_code}
\end{equation}
The \(n_{\mathrm{out}}\) physical qubits of the outer code therefore correspond
to \(n_{\mathrm{out}}\) bosonic data modes. Additional GKP modes serve as
ancillas for inner correction and outer-code stabilizer measurement.

A correction cycle contains three stages. First, inner GKP correction estimates
the displacement of each relevant mode modulo the GKP lattice spacing. This
step produces a corrective displacement, a binary Pauli-frame transition, and a
continuous outcome that quantifies the reliability of the binary decision.
Second, the outer-code stabilizers are measured by coupling GKP data modes to
GKP ancillas according to the supports of the parity checks. Third, the
resulting binary syndrome and GKP-derived likelihoods are processed by a
classical decoder, which returns a recovery or Pauli-frame update.

\begin{figure*}[t]
    \centering
    \includegraphics[
        width=0.95\textwidth,
        keepaspectratio
    ]{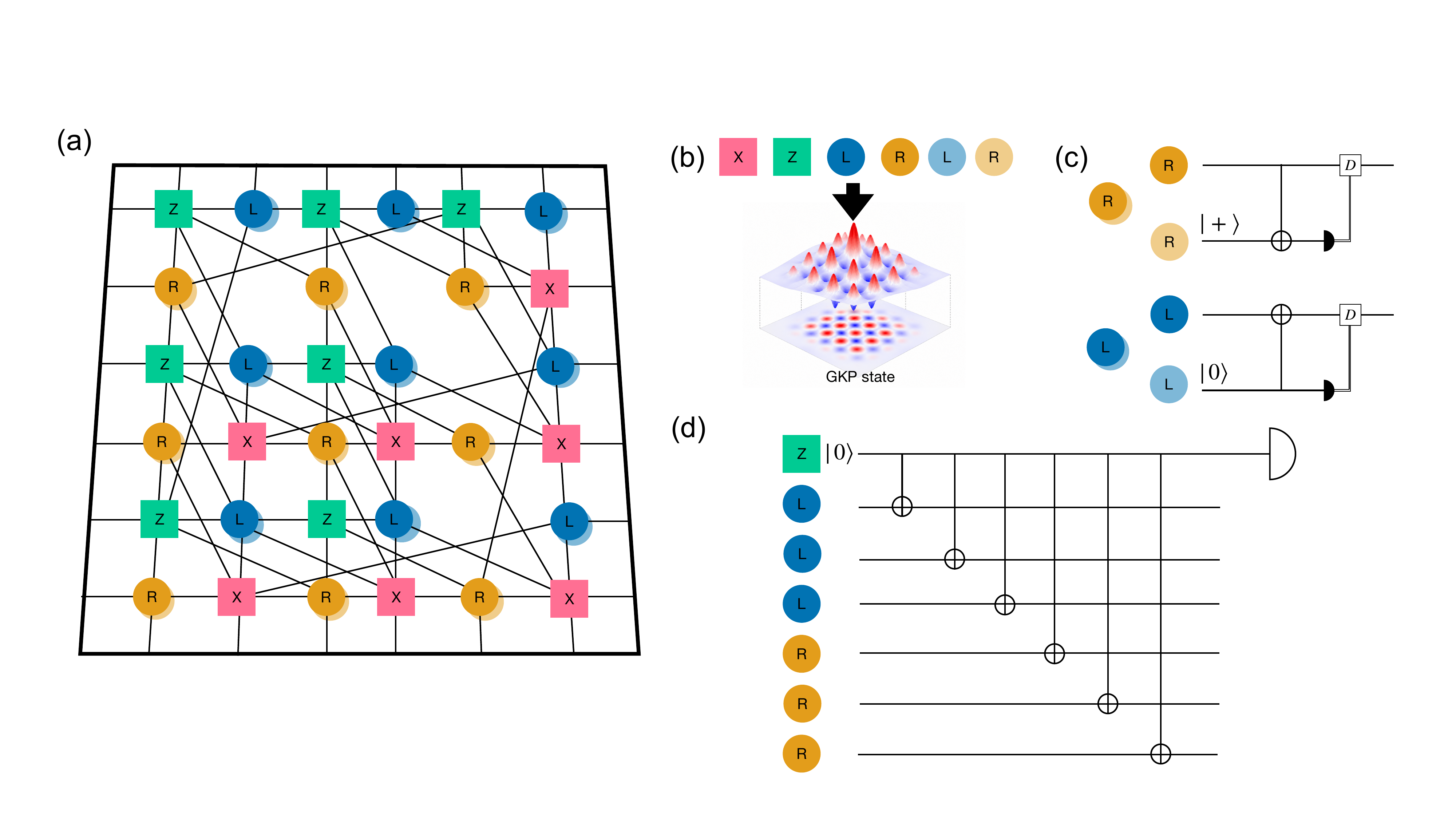}
    \caption{
    Processing stack for GKP-concatenated QEC.
    (a) An \([[18,4,4]]\) BB code is shown as a representative outer code.
    (b) Each physical qubit of the outer code is replaced by a bosonic mode
    encoding a GKP qubit.
    (c) Inner Steane-type GKP correction uses ancillary GKP states, Gaussian
    interactions, homodyne measurements, and classical feedback to estimate
    and suppress displacement errors. The measurement produces both a binary
    correction decision and analog reliability information.
    (d) Example circuit for extracting an outer-code \(Z\)-type stabilizer.
    The participating data modes interact with GKP ancillas through SUM-type
    gates, after which the homodyne outcomes are processed into an outer-code
    syndrome and decoder priors.
    }
    \label{fig:gkp_architecture}
\end{figure*}

Figure~\ref{fig:gkp_architecture} illustrates this organization using a BB code
as a representative outer code. Panel (a) specifies the outer stabilizer-code
structure, while panel (b) replaces each outer-code qubit with a GKP-encoded
oscillator. Panels (c) and (d) show the two operations repeated throughout a
correction cycle: inner GKP correction and outer-code stabilizer extraction.

This separation of functions defines the interface between the bosonic and
outer-code layers. The outer-code family determines the stabilizer constraints
and interaction pattern, while the circuit and noise model determine how
physical displacements are mapped to binary fault variables and decoder
likelihoods.
\subsection{Inner-to-Outer Interface}
\label{subsec:inner_outer_interface}

Inner GKP correction converts the displacement on mode \(i\) into a binary
Pauli-frame transition \(e_i\) and an outcome-dependent reliability. A
displacement assigned to the wrong lattice point produces an effective
\(X\)- or \(Z\)-type fault, while the folded homodyne outcome indicates the
confidence of that assignment. We represent the output of the inner layer as
\begin{equation}
    \bigl(e_i,L_i\bigr),
    \qquad
    L_i=\log\frac{1-p_i}{p_i},
    \label{eq:gkp_layer_output}
\end{equation}
where \(p_i\) is the probability that the GKP rounding decision produces a
logical fault.

In hard-decision decoding, \(p_i\) is determined only by the effective noise
strength at the corresponding location. Modes exposed to the same noise are
therefore assigned the same prior. In analog-informed decoding, \(p_i\) is
conditioned on the folded GKP outcome, so modes close to a lattice boundary are
assigned lower confidence than modes close to a lattice point. This retains
information that would otherwise be discarded by binary rounding
~\cite{fukui2017analog,fukui2018high,vuillot2019quantum}. The GKP conventions
and explicit hard and analog likelihood functions used in this work are given
in Appendix~\ref{app:gkp_likelihoods}.

After this conversion, the oscillators are treated as effective qubits of the
outer code. Its parity-check matrices satisfy
\(H_XH_Z^T=0\pmod 2\) 
and their rows specify the supports of the \(X\)- and \(Z\)-type stabilizers.
Because these effective qubits remain physical oscillator modes, the outer
checks are measured through bosonic interactions rather than direct
discrete-variable gates.

For the representative \(Z\)-type circuit in
Fig.~\ref{fig:gkp_architecture}(d), an ancillary GKP mode interacts with every
data mode in the stabilizer support through SUM or inverse-SUM gates. Homodyne
measurement of the ancilla is then converted into the stabilizer outcome. The
corresponding \(X\)-type circuit uses the complementary quadrature and gate
orientation.

The parity-check matrices therefore determine not only the code space but also
the structure of the physical correction circuit. For an outer code with \(n\)
data modes, \(m\) measured checks, and check weights
\(\{w_a\}_{a=1}^{m}\), one syndrome-extraction round contains approximately
\begin{equation}
    N_{\mathrm{SUM}}
    =
    \sum_{a=1}^{m}w_a
    \label{eq:number_sum_gates}
\end{equation}
SUM-type interactions before repeated measurements, ancilla reuse, or
additional inner-correction steps are included.

These interactions propagate continuous displacements between data and ancilla
modes. A fault on an ancilla can produce back-action on the data, while a data
displacement can alter subsequent syndrome outcomes. The effective outer-code
noise therefore depends on the check weights, data-mode degrees, gate ordering,
schedule depth, ordering of the \(X\)- and \(Z\)-check circuits, frequency of
inner correction, ancilla noise, measurement noise, and idle exposure.

The Tanner graph must also be divided into nonconflicting gate layers because a
data mode or ancilla cannot participate in multiple simultaneous interactions.
Consequently, outer codes with similar block parameters may induce different
circuit depths and different propagated-noise distributions. This dependence is
one reason that code-capacity performance alone does not determine
circuit-level performance.

For resource bookkeeping, we distinguish data modes, GKP ancillas, SUM-type
interactions, homodyne measurements, feedback operations, and schedule layers.
We do not assign platform-specific costs to these components because GKP-state
preparation, storage, connectivity, and measurement costs differ substantially
among optical, microwave, and trapped-ion implementations.

\subsection{Decoding Interface}
\label{subsec:decoding_interface}

The decoder connects the inner GKP layer to the outer stabilizer code. It
combines the binary outer-code syndrome with the reliability information
obtained from GKP correction and returns a recovery or Pauli-frame update.
Hard-decision decoding uses average error probabilities determined by the
effective noise strength, whereas analog-informed decoding assigns
outcome-dependent probabilities using the folded GKP residual. Comparing these
two inputs isolates the benefit of retaining analog information.

For repeated noisy syndrome extraction, the decoder uses the syndrome history
to distinguish data faults from measurement faults. The same interface is
applied to both BB and tricycle codes; the code family enters through its
parity-check matrices and syndrome-extraction circuit. The explicit likelihood
functions, BP--OSD procedure, spacetime construction, and logical-failure
criterion are given in Sec.~\ref{subsec:decoder}.

\subsection{Comparison Basis}
\label{subsec:comparison_basis}

Comparisons among GKP-concatenated codes depend jointly on the outer-code
construction, physical noise model, syndrome-extraction circuit, and use of
analog information in decoding. Table~\ref{tab:gkp_prior_benchmarks}
summarizes these choices and the numerical results reported for representative
schemes, together with the circuit-level crossing estimates obtained in this
work. Because the studies use different noise parameters, circuit assumptions,
and performance criteria, their numerical values are not directly comparable.
A meaningful comparison of outer codes therefore requires them to be evaluated
under the same noise models, circuit assumptions, and decoding procedure, as
done here for the selected BB and tricycle sequences.

\begin{table*}[t]
\centering
\footnotesize
\setlength{\tabcolsep}{4.5pt}
\renewcommand{\arraystretch}{1.08}
\begin{tabular}{
    @{}
    l
    >{\raggedright\arraybackslash}p{0.18\textwidth}
    >{\raggedright\arraybackslash}p{0.18\textwidth}
    >{\centering\arraybackslash}p{0.22\textwidth}
    @{}
}
\hline
\textbf{Scheme}
&
\textbf{Noise model}
&
\textbf{Decoder}
&
\textbf{Reported numerical result}
\\
\hline\hline

Toric--GKP~\cite{vuillot2019quantum}
&
Gaussian displacement with repeated GKP correction
&
GKP likelihood tracking and toric decoding
&
\(\sigma_{\mathrm{th}}\simeq0.243^{a}\)
\\

Surface--GKP~\cite{noh2020faulttolerant,noh2020fault}
&
Finite squeezing and circuit-level noise
&
GKP-weighted MWPM
&
\(11.2\,\mathrm{dB}^{b}\)
\\

Low-overhead surface--GKP~\cite{noh2022lowoverhead}
&
Finite-squeezing gate model
&
Gate-level ML and MWPM
&
\(9.9\,\mathrm{dB}^{c}\)
\\

XZZX--GKP~\cite{zhang2023concatenation}
&
Code-capacity and circuit-level noise
&
Analog MWPM
&
\(\sigma_{\mathrm{th}}\simeq0.67^{d}\);
\(16.5\,\mathrm{dB}^{e}\)
\\

Color--GKP~\cite{zhang2021quantum}
&
Gaussian shifts with noisy measurements
&
MWPM and restriction decoding
&
\(\sigma_{\mathrm{th}}\simeq0.59^{f}\);
\(0.24^{g}\)
\\

Repetition--GKP~\cite{stafford2023biased}
&
Biased GKP code under isotropic Gaussian noise
&
Bias-aware decoding
&
\(\sigma_{\mathrm{th}}\simeq0.599^{h}\)
\\

Lifted-product--GKP~\cite{raveendran2022finite}
&
Gaussian displacement with ideal GKP ancillas
&
Sequential min-sum decoding
&
\(\sigma_{\mathrm{th}}>0.524^{i}\)
\\

BB--GKP~\cite{borah2025faulttolerantdecodingqldpcgkp}
&
Circuit-level GKP noise
&
BP--OSD with real-time soft information
&
Up to \(10^{2}\) FER reduction\(^{j}\)
\\

Photonic qLDPC--GKP~\cite{walshe2025linear}
&
Finite squeezing in a linear-optical architecture
&
MWPM and BP--OSD
&
\(\sim10.9\,\mathrm{dB}\)
(\(\sigma\simeq0.202\))\(^{k}\)
\\

\hline

This work: BB--GKP
&
Code capacity, variance aggregation, and schedule-resolved circuit noise
&
Hard-decision and analog-informed BP--OSD
&
\(\sigma_{\mathrm{cross}}\simeq0.212^{l}\)
\\

This work: tricycle--GKP
&
Code capacity, variance aggregation, and schedule-resolved circuit noise
&
Hard-decision and analog-informed BP--OSD
&
\(\sigma_{\mathrm{cross}}\simeq0.142^{l}\)
\\

\hline
\end{tabular}

\vspace{3pt}
\begin{minipage}{0.98\textwidth}
\scriptsize
\(^{a}\) Data modes, GKP ancillas, and syndrome measurements are noisy.
\(^{b}\) Circuit-level surface--GKP model.
\(^{c}\) Finite-squeezing gate model with optimized decoding.
\(^{d}\) Code-capacity model with rectangular-lattice parameter
\(\lambda=2.1\).
\(^{e}\) Circuit-level model with gate-level maximum-likelihood decoding.
\(^{f}\) Only the data GKP modes are noisy.
\(^{g}\) Data modes and GKP measurements are both noisy.
\(^{h}\) Code-capacity model with optimized rectangular-lattice bias.
\(^{i}\) The LP04 and LP118 families have asymptotic rates \(0.04\) and
\(0.118\), respectively. Analog-informed sequential min-sum decoding produces
crossings beyond the CSS Hamming-bound value \(\sigma=0.524\); the estimate
depends on the code family and decoder configuration.
\(^{j}\) Circuit-level simulations include the \([[144,12,12]]\) BB code
with a depth-six extraction schedule and a lifted-product code. The study
reports finite-code frame-error-rate curves rather than an asymptotic
threshold.
\(^{k}\) Simulations include hyperbolic surface codes and the
\([[72,12,6]]\), \([[90,8,10]]\), \([[144,12,12]]\), and
\([[288,12,18]]\) BB codes. The quoted value is a representative performance
crossover in the hyperbolic-code comparison, not a universal qLDPC threshold.
\(^{l}\) Analog-informed finite-size crossing estimate under the
schedule-resolved circuit model. The hard-decision estimates are \(0.198\)
for the BB sequence and \(0.130\) for the tricycle sequence. The analog
values correspond to equivalent squeezing levels of \(10.46\,\mathrm{dB}\)
and \(13.94\,\mathrm{dB}\), respectively.
\end{minipage}

\caption{
Representative numerical results for GKP-concatenated codes, together with
the circuit-level finite-size crossing estimates obtained in this work.
Values are reported using the conventions of the corresponding studies;
displacement-noise thresholds, squeezing thresholds, performance crossovers,
and finite-size error-rate reductions are not directly comparable.
}
\label{tab:gkp_prior_benchmarks}
\end{table*}

Previous results consistently show that modular GKP outcomes can improve
outer-code decoding, while circuit-level performance depends on check weight,
schedule depth, repeated measurements, and displacement propagation
~\cite{vuillot2019quantum,noh2020fault,noh2022lowoverhead,
raveendran2022finite,borah2025faulttolerantdecodingqldpcgkp}. We therefore hold
the physical noise parameters, circuit conventions, and decoder settings fixed
when comparing the selected BB and tricycle instances.

\subsection{Specialization to BB and Tricycle Codes}
\label{subsec:motivation_bb_tricycle}

Finite-rate qLDPC codes are natural outer-code candidates because their sparse
parity-check matrices support low-weight syndrome extraction and message-passing
decoders, while their nonvanishing rates may reduce asymptotic storage overhead.
BB and tricycle codes provide two structured families with different block
forms, stabilizer profiles, and syndrome-extraction circuits.

Previous GKP--BB studies have established promising behavior in complementary
settings. Selected BB instances have exhibited favorable encoding rates and
competitive thresholds in a passive-linear-optical measurement-based
architecture~\cite{walshe2025linear}. A circuit-level study of the
\([[144,12,12]]\) BB code also found substantial gains from real-time GKP soft
information using a fixed syndrome-extraction and inner-correction schedule
~\cite{borah2025faulttolerantdecodingqldpcgkp}.

These results do not by themselves determine how BB codes compare with another
finite-rate family. The photonic analysis used homogeneous Gaussian noise and a
selected set of BB instances, whereas the circuit-level analysis focused on one
BB code and one fixed correction protocol. Differences in noise, circuit
construction, and decoding assumptions make cross-study threshold comparisons
inconclusive.

We therefore evaluate selected BB and tricycle sequences using common physical
noise parameters, syndrome-generation rules, scheduling conventions, and
decoder settings. The same hard-decision and analog-informed BP--OSD interfaces
are applied to both families. The comparison is performed successively under
code-capacity noise, repeated noisy syndrome measurements, and
schedule-resolved displacement propagation. This construction isolates how the
outer parity-check structure affects finite-size crossing estimates as
additional circuit-level mechanisms are introduced.
\section{Numerical Simulation Framework}
\label{sec:numerical_framework}

To separate the effects of outer-code structure, analog GKP information, and
circuit-level noise propagation, we evaluate selected BB and tricycle instances
within a common numerical framework. For each noise model, both code families
are simulated using the same physical-noise parameters, syndrome-generation
rules, and decoder settings. Hard-decision and analog-informed decoding are
also applied to the same sampled noise realizations, allowing the contribution
of the modular GKP outcomes to be isolated from changes in the physical channel.

The framework contains three levels of circuit detail. The code-capacity model
provides an idealized baseline with independent displacement noise and perfect
syndrome extraction. The variance-aggregation model adds repeated noisy
measurements and incorporates the dependence of accumulated noise on check
weights and data-mode degrees. The schedule-resolved circuit model then
propagates continuous displacements through the ordered syndrome-extraction
circuit, including ancilla, gate, measurement, and idle noise. This section
specifies the code instances, fault-generation procedures, common BP--OSD
interface, logical-failure criterion, and finite-size crossing procedure used
throughout the simulations. The resulting numerical comparisons are presented
in Sec.~\ref{sec:numerical_results}.

\subsection{Outer-Code Instances}
\label{subsec:outer_code_instances}

We consider BB codes~\cite{bravyi2024high} and tricycle
codes~\cite{menon2026magictricycles} as representative finite-rate qLDPC outer codes. These code families were introduced in previous work; our
contribution is their evaluation as outer codes in a common GKP-concatenated
simulation framework. The simulator accesses each code only through binary CSS
parity-check matrices
\begin{equation}
    H_X,H_Z\in\mathbb{F}_2,
    \qquad H_XH_Z^T=0\pmod 2.
    \label{eq:numerical_css_interface}
\end{equation}
Once these matrices are supplied, the same noise sampling, syndrome generation,
decoding, and logical-failure tests are used for both families.

Table~\ref{tab:outer_code_instances} lists the finite-length instances used in
the threshold calculations. Their complete algebraic constructions, generating
polynomials, and matrix checks are given in
Appendix~\ref{app:code_construction}.

\begin{table}[t]
    \centering
    \small
    \setlength{\tabcolsep}{5pt}
    \renewcommand{\arraystretch}{1.12}
    \begin{tabular}{c c c c}
        \hline
        Family & Code parameters & $N$ & $K$ \\
        \hline
        BB & $[[18,4,4]]$       & 18  & 4  \\
        BB & $[[72,12,6]]$      & 72  & 12 \\
        BB & $[[144,12,12]]$    & 144 & 12 \\
        \hline
        Tricycle & $[[48,6,(8,4)]]$   & 48  & 6 \\
        Tricycle & $[[84,6,(12,5)]]$  & 84  & 6 \\
        Tricycle & $[[108,6,(12,6)]]$ & 108 & 6 \\
        \hline
    \end{tabular}
    \caption{Outer-code instances used in the numerical simulations. For a
    tricycle code, the final pair denotes $(D_X,D_Z)$. Detailed generating
    data are provided in Appendix~\ref{app:code_construction}.}
    \label{tab:outer_code_instances}
\end{table}

The two sequences differ in block length, rate, and distance and therefore do
not constitute an asymptotically matched family comparison. Instead, they are
representative finite-length sequences evaluated under identical physical-noise
parameters, circuit conventions, and decoder settings. The reported crossings
should consequently be interpreted as finite-size estimates for these selected
instances rather than universal thresholds of the full code families.

\subsection{Noise Models}
\label{sec:noise_models}

We use three noise models to separate intrinsic code performance from the
effects of repeated syndrome measurements and circuit-level displacement
propagation. The models share the same scan parameter
\(\sigma_{\mathrm{gkp}}\) and the same hard-decision or analog-informed decoder
interface. They differ in the fault locations and circuit information retained
in the simulation, as summarized in Table~\ref{tab:noise_model_summary}.

\begin{table*}[t]
    \centering
    \footnotesize
    \setlength{\tabcolsep}{5pt}
    \renewcommand{\arraystretch}{1.1}
    \begin{tabular}{l c c c c c c}
        \hline
        \textbf{Model}
        &
        \textbf{Rounds}
        &
        \textbf{Meas. faults}
        &
        \textbf{Weight dep.}
        &
        \textbf{Schedule}
        &
        \textbf{Propagation}
        &
        \textbf{Idle noise}
        \\
        \hline\hline

        Code capacity
        &
        No
        &
        No
        &
        No
        &
        No
        &
        No
        &
        No
        \\

        Variance aggregation
        &
        \(T=3\)
        &
        Yes
        &
        Yes
        &
        No
        &
        No
        &
        No
        \\

        Propagated circuit
        &
        \(T=6\)
        &
        Yes
        &
        Yes
        &
        Yes
        &
        Yes
        &
        Yes
        \\

        \hline
    \end{tabular}
    \caption{
    Information retained by the three noise models. Repeated noisy rounds are
    followed by an ideal syndrome readout.
    }
    \label{tab:noise_model_summary}
\end{table*}

The code-capacity model provides an idealized baseline in which independent
Gaussian displacements act only on the data modes and syndrome extraction is
perfect. The variance-aggregation model adds repeated data and measurement
faults and assigns effective noise strengths according to the row and column
weights of the parity-check matrix, but it does not resolve individual gates or
their ordering. The propagated circuit model executes the outer
syndrome-extraction schedule layer by layer and explicitly includes data,
ancilla, gate, measurement, and idle noise together with continuous
displacement propagation through SUM gates.

The three models therefore form a controlled hierarchy. The first transition
introduces repeated noisy measurements and matrix-dependent circuit exposure,
whereas the second introduces schedule-dependent propagation, ancilla
back-action, and idle exposure. Inner GKP correction is represented as an
idealized effective operation in all three models. Within each model,
hard-decision and analog-informed decoding are applied to the same sampled
realization, so their comparison isolates the benefit of retaining analog GKP
information.

\subsubsection{Code-Capacity Model}
\label{subsec:code_capacity_noise}

In the code-capacity setting, independent displacement noise acts only on the
data GKP modes; inner GKP correction and outer syndrome extraction are ideal.
For every data mode $i$, we sample
\begin{equation}
    \xi_{q,i},\xi_{p,i}\sim
    \mathcal{N}(0,\sigma_{\mathrm{gkp}}^2).
    \label{eq:capacity_gaussian_shifts}
\end{equation}
Let $s=\sqrt{\pi}$ be the logical GKP lattice spacing. Ideal GKP correction
folds a displacement into $[-s/2,s/2)$ according to
\begin{equation}
    R_s(\xi)=
    \xi-s\left\lfloor\frac{\xi}{s}+\frac{1}{2}\right\rfloor .
    \label{eq:capacity_folding_function}
\end{equation}
The parity of the removed lattice displacement defines the effective Pauli
faults,
\begin{equation}
    e_i^X=\left\lfloor\frac{\xi_{q,i}}{s}+\frac12\right\rfloor\bmod2,
    \qquad
    e_i^Z=\left\lfloor\frac{\xi_{p,i}}{s}+\frac12\right\rfloor\bmod2.
    \label{eq:capacity_effective_pauli_errors}
\end{equation}
The folded residuals $R_s(\xi_{q,i})$ and $R_s(\xi_{p,i})$ are retained for
analog-informed decoding. The ideal CSS syndromes are
\begin{equation}
    \boldsymbol{s}_Z=H_Z\boldsymbol{e}^X,
    \qquad
    \boldsymbol{s}_X=H_X\boldsymbol{e}^Z
    \pmod 2.
    \label{eq:capacity_ideal_syndromes}
\end{equation}

\subsubsection{Statistical Variance-Aggregation Model}
\label{subsec:variance_aggregation_noise}

As an intermediate model, we retain repeated noisy syndrome measurements and a
spacetime decoding problem but replace individual SUM-gate propagation by
effective Gaussian channels. For either Pauli branch, let
$H\in\mathbb{F}_2^{m\times n}$ be the relevant check matrix. We use $H=H_Z$
for $X$ errors and $H=H_X$ for $Z$ errors. The simulations contain $T=3$
noisy extraction cycles followed by a final ideal syndrome.

Let $\boldsymbol{f}_t$ be newly generated data faults and
$\boldsymbol{\mu}_t$ measurement faults in cycle $t$. The cumulative data error
and measured syndrome are
\begin{equation}
    \boldsymbol{E}_t=\sum_{\tau=0}^{t}\boldsymbol{f}_\tau\pmod2,
    \qquad
    \boldsymbol{y}_t=H\boldsymbol{E}_t+\boldsymbol{\mu}_t\pmod2,
    \label{eq:aggregate_noisy_syndrome}
\end{equation}
with terminal ideal readout
$\boldsymbol{y}_T=H\boldsymbol{E}_{T-1}$.

Define the column degree and row weight of $H$ by
\begin{equation}
    w_j^{\mathrm{col}}=\sum_{a=1}^{m}H_{aj},
    \qquad
    w_a^{\mathrm{row}}=\sum_{j=1}^{n}H_{aj}.
    \label{eq:aggregate_row_column_weights}
\end{equation}
Independent Gaussian contributions are aggregated at the variance level,
\begin{equation}
    \begin{split}
        \sigma_{\mathrm{data},j}^2
        &=(\lambda_{\mathrm{data}}\sigma_{\mathrm{gkp}})^2
        +w_j^{\mathrm{col}}
        (\lambda_{\mathrm{gate}}\sigma_{\mathrm{gkp}})^2,\\
        \sigma_{\mathrm{meas},a}^2
        &=(\lambda_{\mathrm{meas}}\sigma_{\mathrm{gkp}})^2
        +w_a^{\mathrm{row}}
        (\lambda_{\mathrm{gate}}\sigma_{\mathrm{gkp}})^2.
    \end{split}
    \label{eq:aggregate_measurement_variance}
\end{equation}
We set $\lambda_{\mathrm{data}}=1$, $\lambda_{\mathrm{gate}}=1/2$, and
$\lambda_{\mathrm{meas}}=1$.

The syndrome history is converted into detection events,
\begin{align}
    \boldsymbol{d}_0
    &=H\boldsymbol{f}_0+\boldsymbol{\mu}_0,\nonumber\\
    \boldsymbol{d}_t
    &=H\boldsymbol{f}_t+\boldsymbol{\mu}_t+\boldsymbol{\mu}_{t-1},
    &&1\leq t\leq T-1,\label{eq:aggregate_detection_events}\\
    \boldsymbol{d}_T&=\boldsymbol{\mu}_{T-1}.\nonumber
\end{align}
Equivalently, $\boldsymbol{d}=M_H\boldsymbol{u}$ over $\mathbb{F}_2$, where
$\boldsymbol{u}$ contains all data- and measurement-fault variables. This model
captures repeated faults, faulty syndrome outcomes, and dependence on check
weight and data degree. It does not resolve gate ordering, propagated
quadrature correlations, idle periods, or cancellations between SUM and
inverse-SUM gates.

\subsubsection{Full Propagated Circuit-Level Model}
\label{subsec:full_circuit_noise}

The full model propagates the continuous displacement of every data and check
ancilla mode through the scheduled GKP--CSS syndrome-extraction circuit. Each
nonzero entry of $H$ defines a Tanner-graph edge,
\begin{equation}
    \mathcal{E}(H)=\{(a,j)\mid H_{aj}=1\}.
    \label{eq:full_tanner_edges}
\end{equation}
A greedy edge coloring partitions these edges into layers
$\mathcal{E}(H)=\bigsqcup_{\ell=1}^{D_H}\mathcal{L}_\ell$, with no data mode
or ancilla participating in more than one gate in the same layer. Independent
schedules are generated for $H_Z$ and $H_X$, and the $Z$-check layers precede
the $X$-check layers in every cycle.

All physical widths are controlled by $\sigma_{\mathrm{gkp}}$ through
\begin{equation}
    \sigma_\alpha=\lambda_\alpha\sigma_{\mathrm{gkp}},
    \quad
    \alpha\in\{\mathrm{data},\mathrm{anc},\mathrm{gate},
    \mathrm{meas},\mathrm{idle}\}.
    \label{eq:full_noise_strengths}
\end{equation}
We use
\begin{equation}
    (\lambda_{\mathrm{data}},\lambda_{\mathrm{anc}},
    \lambda_{\mathrm{gate}},\lambda_{\mathrm{meas}},
    \lambda_{\mathrm{idle}})
    =(0.25,1,0.15,0.50,0.05).
    \label{eq:full_noise_coefficients}
\end{equation}
Data noise is applied once per cycle; fresh ancillas are prepared in each
cycle; gate noise is applied to both modes after each SUM interaction; idle
noise is applied to inactive data modes in every layer; and measurement noise
is added before homodyne binning.

At the start of each cycle, idealized inner GKP correction converts the current
continuous data shifts into binary Pauli-frame transitions and modular outcomes
using Eqs.~\eqref{eq:capacity_folding_function} and
\eqref{eq:capacity_effective_pauli_errors}. The binary transitions are retained,
whereas the corrected continuous residual is cleared. Continuous shifts
generated during the subsequent outer-code circuit remain until the next inner
correction.

A SUM gate from control $c$ to target $r$ propagates displacements as
\begin{equation}
    \begin{pmatrix}q_c'\\q_r'\\p_c'\\p_r'\end{pmatrix}
    =
    \begin{pmatrix}
        1&0&0&0\\1&1&0&0\\0&0&1&-1\\0&0&0&1
    \end{pmatrix}
    \begin{pmatrix}q_c\\q_r\\p_c\\p_r\end{pmatrix}.
    \label{eq:full_sum_propagation}
\end{equation}
For a $Z$ check, the data mode controls the ancilla target and the ancilla
$q$ quadrature is measured. For an $X$ check, the ancilla controls the data
target and its $p$ quadrature is measured. If
$\boldsymbol{\mu}_{Z,t}$ and $\boldsymbol{\mu}_{X,t}$ denote the binary faults
obtained by rounding the noisy homodyne outcomes, then
\begin{equation}
    \boldsymbol{y}_{Z,t}=H_Z\boldsymbol{E}_t^X+\boldsymbol{\mu}_{Z,t},
    \qquad
    \boldsymbol{y}_{X,t}=H_X\boldsymbol{E}_t^Z+\boldsymbol{\mu}_{X,t}.
    \label{eq:full_noisy_syndrome_records}
\end{equation}

We simulate $T=6$ noisy cycles, followed by a final ideal inner correction and
ideal outer-syndrome readout. This produces $T+1$ data-fault layers and $T$
measurement-fault layers. For either Pauli branch, the complete detection-event
system is
\begin{equation}
    M_H\boldsymbol{x}_H=\boldsymbol{d}_H,
    \qquad
    M_H=\left[I_{T+1}\otimes H\;\middle|\;B_T\otimes I_m\right],
    \label{eq:full_spacetime_system}
\end{equation}
where $B_T\in\mathbb{F}_2^{(T+1)\times T}$ is the open-boundary temporal
incidence matrix with ones on its main and first subdiagonal. Because $T$ is
fixed rather than scaled with distance, the reported crossings are
finite-duration circuit-level estimates for this protocol, not
$T\propto d$ memory-threshold extrapolations.

Circuit propagation produces non-identical marginal distributions for data,
$Z$-check, and $X$-check variables. Decoder priors use effective widths
$\sigma_{\mathrm{prior},v}=\kappa_v\sigma_{\mathrm{gkp}}$, with frozen values
\begin{equation}
    \kappa_{\mathrm{data}}\simeq1.848,
    \qquad
    \kappa_{Z\text{-check}}\simeq1.273,
    \qquad
    \kappa_{X\text{-check}}\simeq6.679.
    \label{eq:full_calibrated_widths}
\end{equation}
These values were fitted on representative BB circuits using recorded modular
outcomes and realized fault labels, frozen before the threshold scans, and
transferred unchanged to all BB and tricycle instances. Thus, the tricycle
results receive no family-specific prior calibration.

\subsection{Decoder}
\label{subsec:decoder}

All three noise models share the same classical decoder. The $X$ and $Z$
branches are decoded independently using belief propagation with
ordered-statistics post-processing (BP--OSD). The code family enters only
through $H_X,H_Z$ or the corresponding spacetime matrix; analog information
changes the variable priors but not the decoding graph or failure criterion.

\subsubsection{Hard and Analog GKP Priors}

For a Gaussian shift of width $\sigma$, the hard-decision GKP error probability
is
\begin{equation}
    p_{\mathrm{hard}}(\sigma)=
    \sum_{k\in\mathbb{Z}}
    \int_{(2k+1/2)s}^{(2k+3/2)s}
    \frac{e^{-\xi^2/(2\sigma^2)}}{\sqrt{2\pi}\sigma}\,d\xi .
    \label{eq:decoder_hard_probability}
\end{equation}
For a folded residual $z\in[-s/2,s/2)$, the analog-informed probability is
\begin{equation}
    p_{\mathrm{ana}}(z;\sigma)=
    \frac{\displaystyle\sum_{k\in\mathbb{Z}}
    e^{-[z-(2k+1)s]^2/(2\sigma^2)}}
    {\displaystyle\sum_{k\in\mathbb{Z}}
    e^{-(z-ks)^2/(2\sigma^2)}}.
    \label{eq:decoder_analog_probability}
\end{equation}
Each variable receives the log-likelihood ratio
\begin{equation}
    \Lambda_i=\log\frac{1-p_i}{p_i},
    \qquad
    p_i=\begin{cases}
        p_{\mathrm{hard}}(\sigma_i),&\text{hard decision},\\
        p_{\mathrm{ana}}(z_i;\sigma_i),&\text{analog informed}.
    \end{cases}
    \label{eq:decoder_input_probabilities}
\end{equation}
Probabilities are clipped to $[10^{-15},1/2-10^{-15}]$. Hard and analog
decoding are applied to the same sampled realization, so their comparison
isolates the value of retaining the modular GKP outcome.

\subsubsection{BP--OSD Inference}

For one decoder branch, let $\mathsf{A}$ be either a spatial check matrix or a
spacetime matrix, let $\boldsymbol{x}$ denote the fault variables, and let
$\boldsymbol{s}$ be the observed syndrome or detection-event vector. The
decoder seeks
\begin{equation}
    \mathsf{A}\hat{\boldsymbol{x}}=\boldsymbol{s}\pmod2.
    \label{eq:decoder_binary_problem}
\end{equation}
BP performs sum-product message passing on the Tanner graph of $\mathsf{A}$
using the LLRs in Eq.~\eqref{eq:decoder_input_probabilities}. It terminates when
the tentative estimate satisfies Eq.~\eqref{eq:decoder_binary_problem}, or
after a fixed maximum number of iterations. Message damping and clipping are
used to control oscillations and numerical overflow.

If BP does not return a syndrome-consistent estimate, OSD orders variables by
their posterior reliability, performs binary Gaussian elimination on the
permuted matrix, and searches low-weight assignments among the least reliable
free variables. Among syndrome-consistent candidates, it selects the one with
minimum reliability cost. This post-processing is particularly useful for
qLDPC Tanner graphs containing short cycles and redundant checks
~\cite{raveendran2022finite,borah2025faulttolerantdecodingqldpcgkp}.

The settings $(I_{\max},\gamma,\Lambda_{\max},w,K)$ are
$(50,0,50,2,20)$ for code capacity, $(50,0.3,30,0,8)$ for variance
aggregation, and $(150,0.5,10,1,12)$ for the full circuit model. Within each
noise model, these settings are fixed across code families, block lengths,
noise strengths, and hard or analog inputs.

\subsection{Logical Failure and Threshold Estimation}
\label{subsec:logical_failure}

Let $\boldsymbol{e}^X,\boldsymbol{e}^Z$ be the accumulated data errors and
$\hat{\boldsymbol{e}}^X,\hat{\boldsymbol{e}}^Z$ the decoder corrections.
Measurement-fault variables inferred in spacetime are discarded after decoding.
The residuals are
\begin{equation}
    \boldsymbol{r}^X=\boldsymbol{e}^X+\hat{\boldsymbol{e}}^X,
    \qquad
    \boldsymbol{r}^Z=\boldsymbol{e}^Z+\hat{\boldsymbol{e}}^Z
    \pmod2.
\end{equation}
A trial succeeds exactly when both residuals are stabilizers,
\begin{equation}
    \begin{aligned}
        H_Z\boldsymbol{r}^X&=0,
        &\boldsymbol{r}^X&\in\operatorname{row}(H_X),\\
        H_X\boldsymbol{r}^Z&=0,
        &\boldsymbol{r}^Z&\in\operatorname{row}(H_Z).
    \end{aligned}
    \label{eq:decoder_logical_success}
\end{equation}
A word failure is recorded if either branch has an unresolved syndrome or a
syndrome-free residual in a nontrivial logical coset. For
$N_{\mathrm{samp}}$ trials, the word logical error rate is
\begin{equation}
    P_{\mathrm{L}}(\sigma_{\mathrm{gkp}})=
    \frac{N_{\mathrm{word\,fail}}}{N_{\mathrm{samp}}}.
    \label{eq:word_error_estimator}
\end{equation}
For each code family and decoder input, the threshold is estimated from the
finite-size crossings of $P_{\mathrm{L}}$ as a function of
$\sigma_{\mathrm{gkp}}$. Because only three non-identically scaled instances
are available in each family, we report these values as central finite-size
crossing estimates rather than asymptotic critical points.

\section{Numerical Results}
\label{sec:numerical_results}
We now present the numerical results for GKP-concatenated BB and tricycle codes under the three noise models introduced in Sec.~\ref{sec:noise_models}. For each model, we compare hard-decision and analog-informed BP--OSD decoding and estimate the corresponding thresholds from the finite-size crossings of the word logical error rates. This comparison allows us to assess both the benefit of retaining analog GKP information and the effect of progressively incorporating repeated syndrome measurements and circuit-level noise propagation. 

\subsection{Code-Capacity Results}
\label{subsec:capacity_results}

\begin{figure*}[t]
    \centering
    \includegraphics[width=\textwidth]
    {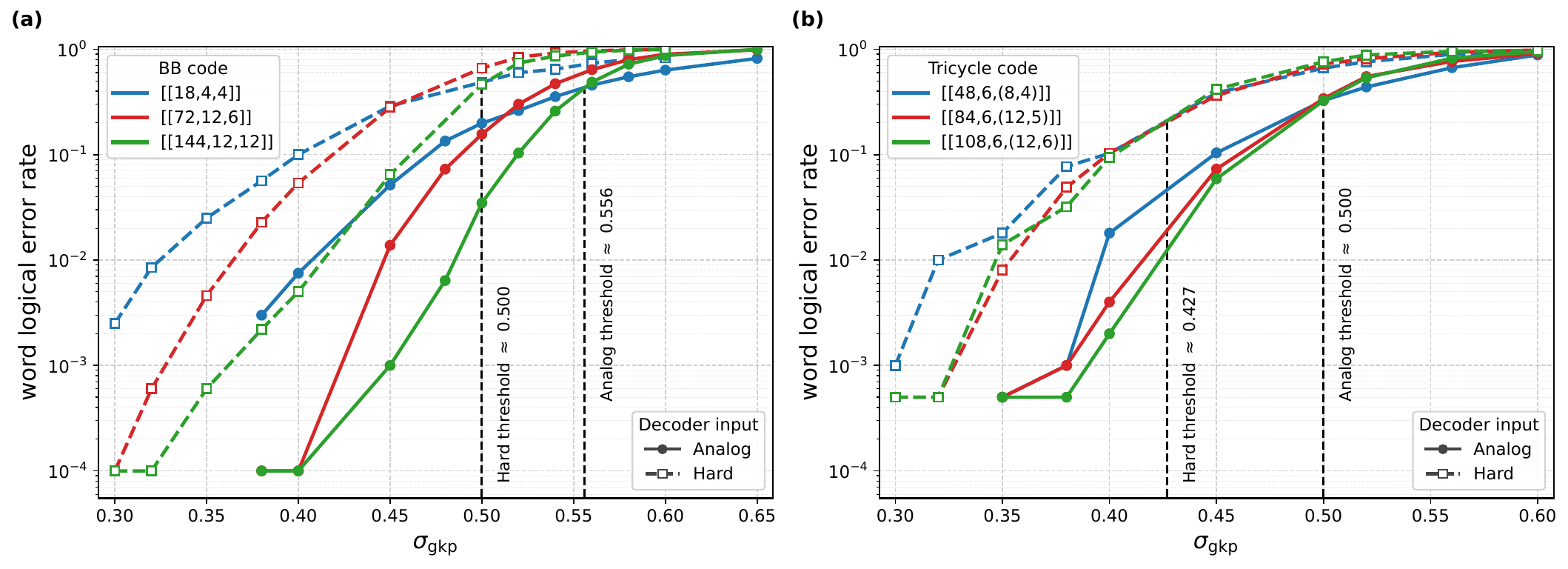}
    \caption{Code-capacity word logical error rates for (a) GKP--BB and
    (b) GKP--tricycle concatenation. Solid curves with filled circles use
    analog-informed priors; dashed curves with open squares use hard-decision
    priors. The finite-size crossing estimates are
    $(\sigma_{\mathrm{th}}^{\mathrm{ana}},
    \sigma_{\mathrm{th}}^{\mathrm{hard}})=(0.556,0.500)$ for BB codes and
    $(0.500,0.427)$ for tricycle codes.}
    \label{fig:capacity_thresholds}
\end{figure*}

Figure~\ref{fig:capacity_thresholds} shows the code-capacity scaling. Analog
information raises the central crossing estimate from $0.500$ to $0.556$ for
BB codes and from $0.427$ to $0.500$ for tricycle codes. Since the sampled
channel and binary syndrome are otherwise unchanged, these shifts directly
quantify the benefit of retaining mode-resolved GKP reliability information.
Under the common simulation assumptions, the selected BB sequence has the
higher crossing for both decoder inputs.

\subsection{Statistical Variance-Aggregation Results}
\label{subsec:aggregate_results}

\begin{figure*}[t]
    \centering
    \includegraphics[width=\textwidth]
    {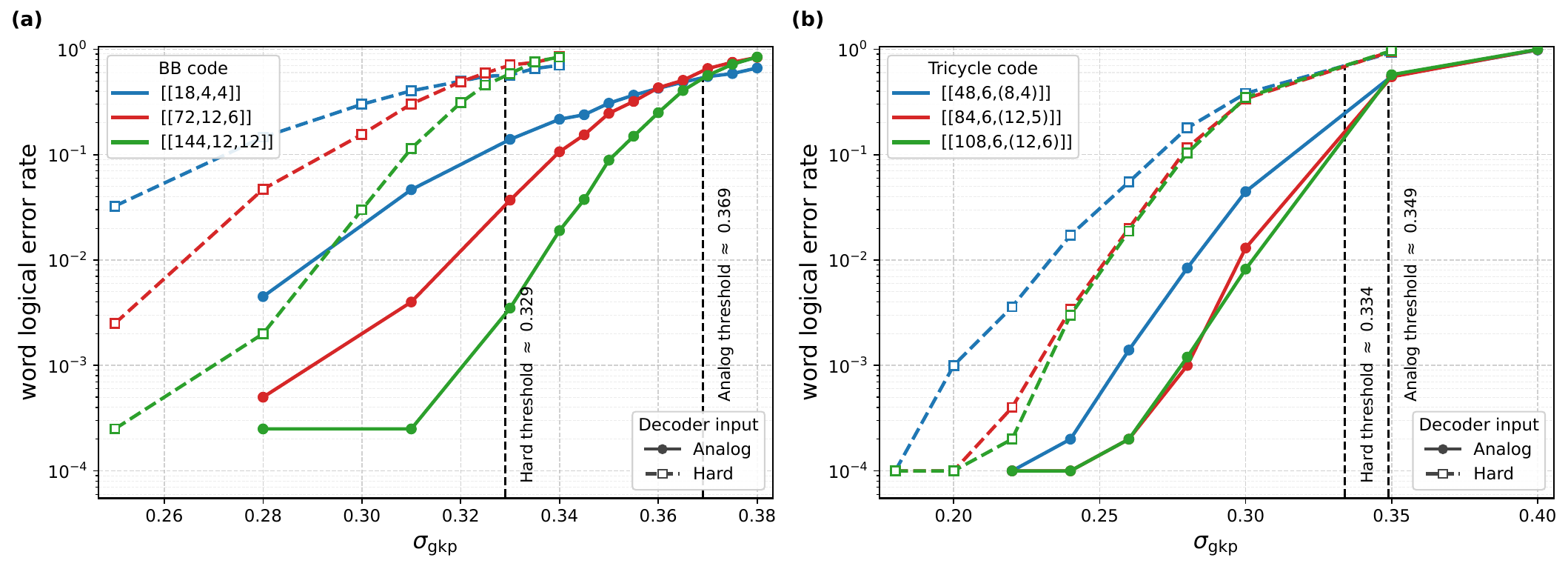}
    \caption{Word logical error rates under the variance-aggregation model for
    (a) GKP--BB and (b) GKP--tricycle concatenation. The simulations use $T=3$
    noisy rounds followed by an ideal syndrome. The crossing estimates are
    $(0.369,0.329)$ for analog and hard BB decoding and $(0.349,0.334)$ for
    analog and hard tricycle decoding, respectively.}
    \label{fig:aggregate_thresholds}
\end{figure*}

As shown in Fig.~\ref{fig:aggregate_thresholds}, repeated data faults and noisy
syndrome measurements lower all crossings relative to code capacity. Analog
priors raise the central estimate from $0.329$ to $0.369$ for BB codes and from
$0.334$ to $0.349$ for tricycle codes. The analog-informed BB estimate is
higher. The hard-decision estimates are close and their finite-size crossing
ranges overlap, so the small reversed ordering is not interpreted as a
significant family-level difference.

\subsection{Full Circuit-Level Results}
\label{subsec:full_circuit_results}

\begin{figure*}[t]
    \centering
    \includegraphics[width=\textwidth]
    {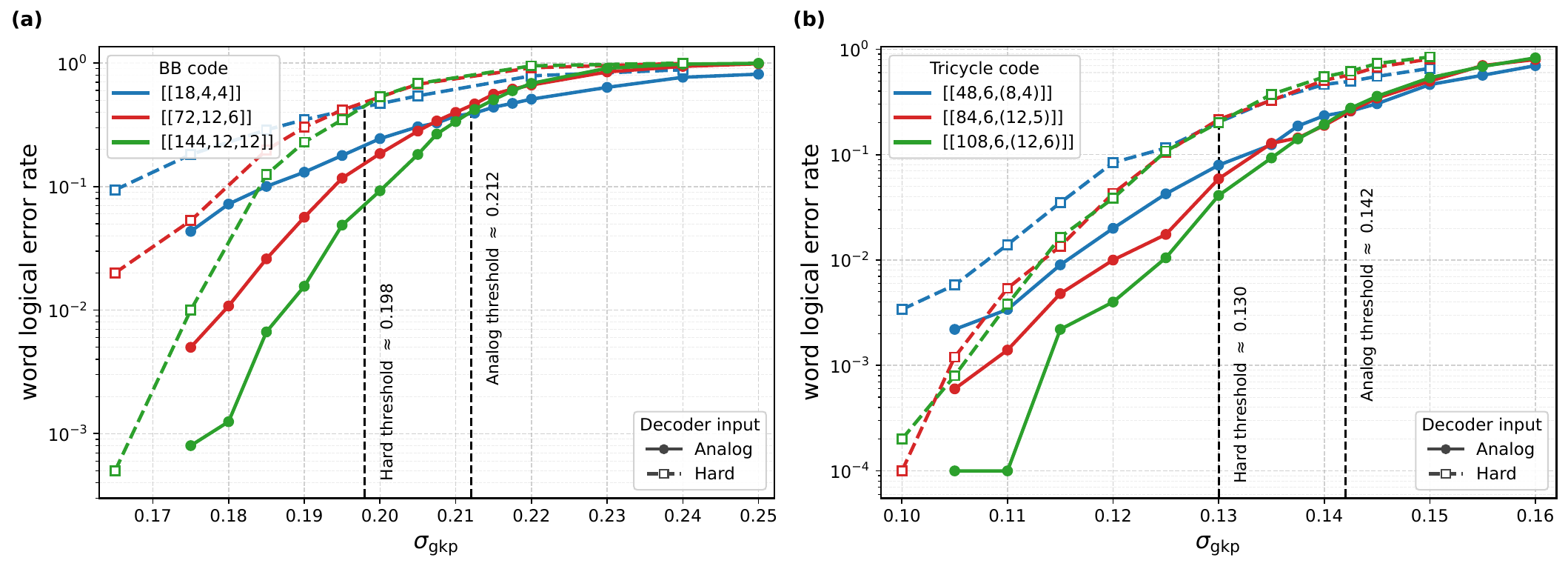}
    \caption{Word logical error rates under the full propagated circuit model
    for (a) GKP--BB and (b) GKP--tricycle concatenation. The finite-size
    crossing estimates are $(0.212,0.198)$ for analog and hard BB decoding and
    $(0.142,0.130)$ for analog and hard tricycle decoding, respectively.}
    \label{fig:full_circuit_thresholds}
\end{figure*}

Figure~\ref{fig:full_circuit_thresholds} presents the results for the frozen
circuit protocol. Analog information raises the central crossing from $0.198$
to $0.212$ for BB codes and from $0.130$ to $0.142$ for tricycle codes. The BB
sequence has the higher crossing for both decoder inputs, and the separation is
larger than in the variance-aggregation model. For the selected instances, the
BB circuits are therefore more robust to schedule-resolved propagation,
measurement asymmetry, and idle exposure.

\subsection{Comparison Across Noise Models}
\label{subsec:cross_model_comparison}

\begin{table}[t]
    \centering
    \small
    \setlength{\tabcolsep}{4pt}
    \renewcommand{\arraystretch}{1.12}
    \begin{tabular}{l c c c c}
        \hline
        & \multicolumn{2}{c}{BB} & \multicolumn{2}{c}{Tricycle} \\
        Model & Hard & Analog & Hard & Analog \\
        \hline
        Code capacity       & 0.500 & 0.556 & 0.427 & 0.500 \\
        Variance aggregation& 0.329 & 0.369 & 0.334 & 0.349 \\
        Full circuit        & 0.198 & 0.212 & 0.130 & 0.142 \\
        \hline
    \end{tabular}
    \caption{Central finite-size crossing estimates
    $\sigma_{\mathrm{th}}$ for the selected BB and tricycle sequences.}
    \label{tab:threshold_summary}
\end{table}

Table~\ref{tab:threshold_summary} summarizes the three comparisons. Increasing
circuit detail systematically lowers the estimated crossings because more
fault locations and propagation mechanisms are included. Analog-informed
decoding improves the central estimate in every model and for both code
families. The selected BB sequence has the larger analog-informed estimate
throughout and also leads under hard decoding in the code-capacity and full
circuit models.

These findings support two conclusions. First, modular GKP outcomes remain
useful after repeated syndrome extraction and circuit-level propagation, not
only in an ideal code-capacity setting. Second, outer-code performance cannot
be inferred from rate and nominal distance alone: check structure, circuit
scheduling, and propagated noise materially affect the observed crossing. The
results establish a controlled finite-length benchmark for the studied code
instances; broader family-level conclusions will require longer sequences and
asymptotic finite-size scaling.

\section{Discussion}

Two trends persist across all three noise models. First, retaining the modular GKP outcomes increases the finite-size crossing estimate relative to hard-decision decoding for both outer-code families, showing that analog information remains useful as the problem develops from independent GKP rounding errors into nonuniform spatial reliabilities and repeated spacetime syndromes. Second, the studied BB sequence has the higher analog-informed crossing in every model. The hard-decision results show the same ordering under code capacity and full circuit noise, while the small reversed ordering in the variance-aggregation model lies within the overlapping crossing ranges and is not statistically resolved. These results favor the BB instances considered here, although they do not imply a universal ordering of all BB and tricycle constructions.

The full circuit results further show that an outer code cannot be assessed from its block parameters alone. Stabilizer weights, Tanner-graph structure, gate scheduling, noise propagation, and decoder response jointly determine how physical displacement noise is converted into data and measurement faults. The three models should therefore be regarded as complementary rather than interchangeable benchmarks: code capacity isolates the intrinsic compatibility between GKP information and the outer-code Tanner graph, variance aggregation adds matrix-dependent circuit exposure at relatively low cost, and full propagation resolves the ordered syndrome-extraction circuit. The first two models can consequently screen code instances and identify relevant noise windows before more expensive circuit-level simulations, whereas quantitative fault-tolerance claims should ultimately be verified under the full circuit model.

Several extensions are possible within the same $(H_X,H_Z)$ interface. Optimized conflict-free schedules, alternative check orderings, and broader CSS-code searches may improve circuit-level performance. Analytic prior propagation and correlation-aware inference could also exploit more of the structure generated by the circuit than the present local-marginal BP--OSD model~\cite{raveendran2022finite,borah2025faulttolerantdecodingqldpcgkp,roy2025decoding}. Hardware-specific resource accounting will eventually be required to determine whether reduced outer-code overhead compensates for bosonic state preparation, repeated correction, homodyne detection, and feedback.

The framework may also be extended to quantum communication and repeater architectures, where displacement noise must be treated together with loss and erasure processes~\cite{rozpdek2021quantumrepeater}. Incorporating loss-induced logical channels and mixed fault types into the matrix-based interface would provide a common route for comparing GKP--qLDPC constructions in both computing and communication settings~\cite{harris2025loss}.

A closely related but complementary benchmark was reported for a passive-linear-optical GKP architecture based on foliated graph states, homogeneous Gaussian noise on the input resource states, and correlation-aware inner decoding followed by BP--OSD~\cite{walshe2025linear}. That work compares BB and hyperbolic surface codes, whereas our study compares BB and tricycle codes through a common $(H_X,H_Z)$ interface and a three-level noise hierarchy that culminates in schedule-resolved circuit simulation. Despite these differences, its BB threshold near $10.6\,\mathrm{dB}$ corresponds to $\sigma_{\mathrm{gkp}}\simeq0.209$ under the same vacuum-noise convention, close to our full circuit-level analog estimate of $\sigma_{\mathrm{th}}^{\mathrm{BB}}\simeq0.212$, or $10.46\,\mathrm{dB}$. This proximity does not imply equivalence of the two noise models, but provides an independent consistency check on the physical threshold scale and analog-decoding methodology. Together with the stable qualitative trends across our hierarchy, it supports using the code-capacity and variance-aggregation models as efficient preliminary tools for ranking new CSS and qLDPC codes and locating their relevant scan regions, followed by full circuit-level validation.

\section{Conclusion}

We have presented a unified numerical framework for evaluating GKP concatenation with CSS qLDPC outer codes. By representing the outer code through its parity-check matrices \((H_X,H_Z)\), the same simulation and BP--OSD decoding pipeline can be applied to different code constructions. We used this framework to compare selected BB and tricycle code sequences across three levels of noise modeling: code capacity, statistical variance aggregation with repeated syndrome measurements, and schedule-resolved circuit-level displacement propagation.

Analog-informed decoding consistently produced higher finite-size crossing estimates than hard-decision decoding for both code families and under all three noise models. In the schedule-resolved circuit model, the analog and hard-decision crossings are \(0.212\) and \(0.198\), respectively, for the selected BB sequence, and \(0.142\) and \(0.130\) for the selected tricycle sequence. These results demonstrate that the reliability information retained by GKP measurements remains useful after repeated noisy syndrome extraction and circuit-level propagation. Under the common assumptions used here, the studied BB sequence also exhibits higher crossing estimates than the studied tricycle sequence.

The comparison shows that the performance of a GKP-concatenated qLDPC code is determined not only by its block parameters, but also by its parity-check structure, syndrome-extraction schedule, displacement propagation, and decoder priors. The reported values should therefore be interpreted as finite-length crossing estimates for the particular code sequences, circuits, and noise models studied here, rather than universal thresholds for the two code families.

More broadly, the framework provides a reproducible route from inexpensive code-capacity screening to more detailed circuit-level evaluation of GKP--CSS concatenations. It can be extended to additional qLDPC constructions, optimized extraction schedules, correlation-aware decoding, and hardware-specific noise and resource models. Such extensions will be necessary to determine whether the finite-size performance observed here translates into a practical reduction in the resources required for bosonic fault-tolerant quantum computation.

\bibliographystyle{apsrev4-2}
\bibliography{sample}

\appendix

\section{GKP Conventions and Analog Likelihoods}
\label{app:gkp_likelihoods}

This appendix specifies the GKP conventions and likelihood functions used in
the architecture and numerical simulations. We take
\begin{equation}
    [\hat q,\hat p]=i
\end{equation}
and use the square GKP lattice with logical spacing
\begin{equation}
    s=\sqrt{\pi}.
    \label{eq:app_gkp_spacing}
\end{equation}

\subsection{Stabilizers and Logical Operators}

The square-lattice GKP stabilizers are
\begin{equation}
    S_q
    =
    \exp\left(i2\sqrt{\pi}\,\hat q\right),
    \qquad
    S_p
    =
    \exp\left(-i2\sqrt{\pi}\,\hat p\right).
    \label{eq:app_gkp_stabilizers}
\end{equation}
The logical Pauli operators are
\begin{equation}
    \overline{Z}
    =
    \exp\left(i\sqrt{\pi}\,\hat q\right),
    \qquad
    \overline{X}
    =
    \exp\left(-i\sqrt{\pi}\,\hat p\right).
    \label{eq:app_gkp_logical_operators}
\end{equation}
The periodic lattice structure permits displacement errors to be measured
modulo \(s\) without directly measuring the encoded logical state.

\subsection{Gaussian Displacement and GKP Rounding}

For a displacement-noise width \(\sigma\), the two quadratures are shifted
according to
\begin{equation}
    q\longrightarrow q+\xi_q,
    \qquad
    p\longrightarrow p+\xi_p,
    \qquad
    \xi_q,\xi_p\sim\mathcal{N}(0,\sigma^2).
    \label{eq:app_gaussian_displacement}
\end{equation}
GKP correction folds a displacement into the fundamental interval
\([-s/2,s/2)\) using
\begin{equation}
    R_s(\xi)
    =
    \xi
    -
    s\left\lfloor
        \frac{\xi}{s}+\frac{1}{2}
    \right\rfloor .
    \label{eq:app_gkp_folding}
\end{equation}
The integer removed during folding is
\begin{equation}
    n(\xi)
    =
    \left\lfloor
        \frac{\xi}{s}+\frac{1}{2}
    \right\rfloor .
    \label{eq:app_gkp_lattice_index}
\end{equation}
Its parity determines the effective Pauli transition,
\begin{equation}
    e(\xi)=n(\xi)\bmod 2.
    \label{eq:app_gkp_binary_fault}
\end{equation}
For the convention used here, \(q\)-quadrature shifts generate effective
\(X\)-type faults and \(p\)-quadrature shifts generate effective \(Z\)-type
faults.

A displacement assigned to an even lattice index is treated as no logical
transition, whereas an odd index produces a logical Pauli transition. The
folded value
\begin{equation}
    z=R_s(\xi)
\end{equation}
is retained as the analog outcome.

\subsection{Hard-Decision Error Probability}

If only the displacement width is retained, the effective GKP error probability
is obtained by integrating the Gaussian distribution over all odd lattice
cells:
\begin{equation}
    p_{\mathrm{hard}}(\sigma)
    =
    \sum_{k\in\mathbb{Z}}
    \int_{(2k+1/2)s}^{(2k+3/2)s}
    \frac{1}{\sqrt{2\pi}\sigma}
    \exp\left(
        -\frac{\xi^2}{2\sigma^2}
    \right)
    d\xi .
    \label{eq:app_gkp_hard_probability}
\end{equation}
All variables with the same effective width receive the log-likelihood ratio
\begin{equation}
    L_{\mathrm{hard}}(\sigma)
    =
    \log
    \frac{1-p_{\mathrm{hard}}(\sigma)}
         {p_{\mathrm{hard}}(\sigma)}.
    \label{eq:app_gkp_hard_llr}
\end{equation}
This prior retains the average error rate but discards the measured location
within the GKP cell.

\subsection{Analog-Informed Error Probability}

Conditioned on a folded outcome \(z\in[-s/2,s/2)\), the probability of an odd
lattice shift is
\begin{equation}
    p_{\mathrm{ana}}(z;\sigma)
    =
    \frac{
        \displaystyle
        \sum_{k\in\mathbb{Z}}
        \exp\left[
            -\frac{\left(z-(2k+1)s\right)^2}{2\sigma^2}
        \right]
    }{
        \displaystyle
        \sum_{k\in\mathbb{Z}}
        \exp\left[
            -\frac{\left(z-ks\right)^2}{2\sigma^2}
        \right]
    }.
    \label{eq:app_gkp_analog_probability}
\end{equation}
The corresponding likelihood is
\begin{equation}
    L_{\mathrm{ana}}(z;\sigma)
    =
    \log
    \frac{1-p_{\mathrm{ana}}(z;\sigma)}
         {p_{\mathrm{ana}}(z;\sigma)}.
    \label{eq:app_gkp_analog_llr}
\end{equation}

Near the center of a GKP cell,
\(p_{\mathrm{ana}}(z;\sigma)\) is small and the magnitude of the likelihood is
large. Near a decision boundary, the probabilities of the adjacent even and
odd lattice assignments become comparable, and the likelihood approaches zero.
Analog-informed decoding therefore distinguishes reliable and unreliable
rounding decisions even when they occur under the same physical noise strength.

In numerical evaluation, the sums in
Eq.~\eqref{eq:app_gkp_analog_probability} are truncated once the omitted
Gaussian terms are negligible at machine precision. Probabilities are clipped
away from zero and \(1/2\) before conversion to likelihoods to avoid numerical
overflow.

\subsection{Relation to Squeezing}

Under the convention that the vacuum quadrature variance is \(1/2\), a Gaussian
displacement width \(\sigma\) is converted to an equivalent squeezing level by
\begin{equation}
    s_{\mathrm{dB}}
    =
    -10\log_{10}\left(2\sigma^2\right).
    \label{eq:app_sigma_to_db}
\end{equation}
A larger displacement threshold therefore corresponds to a lower required
squeezing level. Thresholds expressed in \(\sigma\) or in decibels are
equivalent only when the same quadrature normalization and finite-energy noise
convention are used.

\section{Construction of the BB and Tricycle Code Instances}
\label{app:code_construction}

This appendix gives the algebraic constructions used to generate the BB and
tricycle parity-check matrices in the numerical study. Both families were
introduced previously~\cite{bravyi2024high,menon2026magictricycles}; the details
are included here to identify the simulated instances and make the calculations
reproducible. For every generated matrix pair, we verify CSS commutation,
binary ranks, the number of encoded qubits, and the relevant LDPC row and
column weights.

\subsection{Bivariate Bicycle Codes}
\label{app:bb_construction}

Let $S_\ell$ and $S_m$ be cyclic shift matrices of sizes $\ell$ and $m$, and
define
\begin{equation}
    x=S_\ell\otimes I_m,
    \qquad
    y=I_\ell\otimes S_m.
    \label{eq:app_bb_xy_generators}
\end{equation}
They obey $x^\ell=y^m=I$ and $xy=yx$, providing a matrix representation of
$\mathbb{F}_2[x,y]/(x^\ell-1,y^m-1)$. For the weight-six family used here,
\begin{equation}
    A=x^{a_1}+y^{a_2}+y^{a_3},
    \qquad
    B=y^{b_1}+x^{b_2}+x^{b_3},
    \label{eq:app_bb_AB_polynomials}
\end{equation}
where addition is over $\mathbb{F}_2$. The CSS matrices are
\begin{equation}
    H_X=\begin{pmatrix}A&B\end{pmatrix},
    \qquad
    H_Z=\begin{pmatrix}B^T&A^T\end{pmatrix}.
    \label{eq:app_bb_Hx_Hz}
\end{equation}
Because $A$ and $B$ commute,
\begin{equation}
    H_XH_Z^T=AB+BA=0\pmod2.
\end{equation}

The block length is $N=2\ell m$, with two registers of $\ell m$ qubits. In the
absence of polynomial-term cancellations, every stabilizer has weight six and
every data qubit participates in three checks of each Pauli type. The number of
encoded qubits is evaluated as
\begin{equation}
    K=N-\operatorname{rank}_{\mathbb{F}_2}(H_X)
      -\operatorname{rank}_{\mathbb{F}_2}(H_Z).
\end{equation}
The distances are taken from the established finite-length instances, while
commutation, ranks, and LDPC weights are checked independently.

\begin{table}[ht]
    \centering
    \small
    \setlength{\tabcolsep}{3.5pt}
    \begin{tabular}{c c c c c}
        \hline
        Code & $(\ell,m)$ & $(a_1,a_2,a_3)$
        & $(b_1,b_2,b_3)$ & $N$ \\
        \hline
        $[[18,4,4]]$    & $(3,3)$  & $(1,0,2)$ & $(1,0,2)$ & 18 \\
        $[[72,12,6]]$   & $(6,6)$  & $(3,1,2)$ & $(3,1,2)$ & 72 \\
        $[[144,12,12]]$ & $(12,6)$ & $(3,1,2)$ & $(3,1,2)$ & 144 \\
        \hline
    \end{tabular}
    \caption{BB generating parameters used to construct the parity-check
    matrices in the simulations.}
    \label{tab:app_bb_parameters}
\end{table}

\subsection{Tricycle Codes}
\label{app:tricycle_construction}

Tricycle codes generalize bicycle-type constructions to three homological
dimensions~\cite{menon2026magictricycles}. Let
\begin{equation}
    G=\mathbb{Z}_\ell\times\mathbb{Z}_m\times\mathbb{Z}_n,
    \qquad |G|=\ell mn,
\end{equation}
with generators $x,y,z$. The regular representation maps a monomial to
\begin{equation}
    x^iy^jz^k\longmapsto
    S_\ell^i\otimes S_m^j\otimes S_n^k.
    \label{eq:app_tricycle_regular_representation}
\end{equation}
Binary sums of monomials define sparse $|G|\times|G|$ matrices. Three
group-algebra elements $a,b,c$ define commuting matrices $A,B,C$ through this
map. The CSS checks are
\begin{equation}
    H_X=\begin{pmatrix}A^T&B^T&C^T\end{pmatrix},
    \label{eq:app_tricycle_Hx}
\end{equation}
and
\begin{equation}
    H_Z=
    \begin{pmatrix}
        C&0&A\\
        0&C&B\\
        B&A&0
    \end{pmatrix}.
    \label{eq:app_tricycle_Hz}
\end{equation}
Pairwise terms cancel over $\mathbb{F}_2$, giving
$H_XH_Z^T=0$. The block length and dimension are
\begin{equation}
    N=3\ell mn,
    \qquad
    K=N-\operatorname{rank}_{\mathbb{F}_2}(H_X)
      -\operatorname{rank}_{\mathbb{F}_2}(H_Z).
\end{equation}

The construction also provides the metacheck matrix
\begin{equation}
    H_{\mathrm{meta}}=\begin{pmatrix}B&A&C\end{pmatrix},
    \qquad
    H_{\mathrm{meta}}H_Z=0\pmod2.
    \label{eq:app_tricycle_metacheck}
\end{equation}
In this work, the metacheck relation is used only as an algebraic consistency
test; the simulations use the same $H_X,H_Z$ interface as for BB codes.

\begin{table}[t]
    \centering
    \small
    \setlength{\tabcolsep}{2.5pt}
    \begin{tabular}{c c c l}
        \hline
        Code & $(\ell,m,n)$ & Element & Polynomial \\
        \hline
        $[[48,6,(8,4)]]$ & $(2,2,4)$
            & $a$ & $y+z+xz+xyz^2$ \\
            & & $b$ & $yz^2+yz^3$ \\
            & & $c$ & $y+xyz$ \\
        \hline
        $[[84,6,(12,5)]]$ & $(2,2,7)$
            & $a$ & $y+z+xz+xyz^2$ \\
            & & $b$ & $z^3+xz^4$ \\
            & & $c$ & $y+yz^4$ \\
        \hline
        $[[108,6,(12,6)]]$ & $(3,3,4)$
            & $a$ & $x+z^2+yz+x^2yz^3$ \\
            & & $b$ & $y^2z+x^2yz^3$ \\
            & & $c$ & $x^2+x^2yz^2$ \\
        \hline
    \end{tabular}
    \caption{Tricycle generating polynomials used in the simulations. Each
    monomial is mapped to a binary permutation matrix according to
    Eq.~\eqref{eq:app_tricycle_regular_representation}.}
    \label{tab:app_tricycle_parameters}
\end{table}

\end{document}